\documentclass[reprint, amsmath,amssymb,aps,prb, superscriptaddress]{revtex4-2}
\usepackage{graphicx}
\usepackage{dcolumn}
\usepackage{bm}

\usepackage{xcolor}
\usepackage{longtable}
\usepackage{colortbl}%
  
\usepackage{booktabs,tabularx}
\usepackage{dcolumn}
\usepackage{hyperref}

\hypersetup{
    colorlinks = true,
    citecolor = {blue},
    linkcolor = {blue},
    urlcolor  = {blue},
}

\usepackage{array}
\newcolumntype{P}[1]{>{\centering\arraybackslash}p{#1}}
\newcolumntype{M}[1]{>{\centering\arraybackslash}m{#1}}

\begin{document}

\

\

\title{Magnetoelastic properties in the high-temperature magnetic phase of the skyrmion compound GdRu$_2$Si$_2$}

\author{J. Sourd}
\affiliation{Hochfeld-Magnetlabor Dresden (HLD-EMFL) and Würzburg-Dresden Cluster of Excellence ct.qmat,
Helmholtz-Zentrum Dresden-Rossendorf, 01328 Dresden, Germany}
\author{D. A. Mayoh}
\affiliation{Department of Physics, University of Warwick, Coventry CV4 7AL, United Kingdom}
\author{G. Balakrishnan}
\affiliation{Department of Physics, University of Warwick, Coventry CV4 7AL, United Kingdom}
\author{M. Uhlarz}
\affiliation{Hochfeld-Magnetlabor Dresden (HLD-EMFL) and Würzburg-Dresden Cluster of Excellence ct.qmat,
Helmholtz-Zentrum Dresden-Rossendorf, 01328 Dresden, Germany}
\author{J. Wosnitza}
\affiliation{Hochfeld-Magnetlabor Dresden (HLD-EMFL) and Würzburg-Dresden Cluster of Excellence ct.qmat,
 Helmholtz-Zentrum Dresden-Rossendorf, 01328 Dresden, Germany}
\affiliation{Institut für Festkörper- und Materialphysik, TU Dresden, 01062 Dresden, Germany}
\author{S. Zherlitsyn}
\affiliation{Hochfeld-Magnetlabor Dresden (HLD-EMFL) and Würzburg-Dresden Cluster of Excellence ct.qmat,
Helmholtz-Zentrum Dresden-Rossendorf, 01328 Dresden, Germany}

\date{\today}\begin{abstract}
We investigated the magnetoelastic properties of a GdRu$_2$Si$_2$ single crystal under a magnetic field applied along the crystallographic [001] and [110] directions. We report a series of strong anomalies in the sound velocity that is consistent with the complex phase diagram reported previously for this compound. In particular, in our study we focus on the recently identified magnetic phase in the high-temperature region. We show that while this phase is easily destroyed for magnetic fields applied along [001], it is rather stable for fields along [110]. Furthermore, we introduce a Landau theory and a microscopic toy model describing the elastic response at zero field. We reproduce qualitatively the observed anomalies for different acoustic modes, which allows us to propose a magnetic structure for this new high-temperature phase.

% Furthermore, the different transitions are better resolved with acoustic modes of specific symmetry. We introduce a microscopic toy model where the spin-lattice interactions are introduced through a non-local Kondo coupling between the magnetic moments generated by Gd $4f$ electrons, and the conduction Ru $4d$ electrons. This model permits to reproduce the set of anomalies observed for phase V, and allows us to propose a magnetic structure for this phase.
\end{abstract}

\maketitle

\section{\label{sec:level1}Introduction}

The magnetism of metallic systems is very rich and complex due to the important role of itinerant electrons, which carry the spin density in a complex and self-interacting fashion \cite{brando2016metallic}. In the rare-earth metals, however, the magnetic properties are often dominated by localized $4f$ electrons that host a large magnetic moment. From this perspective, one can separate two types of degrees of freedom: localized $4f$ magnetic moments that can behave quasi-classically on the one hand and conduction electrons that generate the magnetic interactions between the localized moments on the other hand \cite{szilva2023quantitative}. This permits integrating out the contribution from the conduction electrons, and within second-order perturbation theory one obtains the Ruderman-Kittel-Kasuya-Yosida (RKKY) interaction between pairs of localized moments \cite{ruderman1954indirect,kasuya1956theory,yosida1957magnetic}. In this way, the RKKY scheme reduces the physics of interacting localized and itinerant electrons into a Heisenberg-like model of classical spins on a lattice, much easier to deal with.

RKKY interactions are long ranged and oscillatory, often driving incommensurate or spiral magnetic order with propagation vectors determined by the Fermi-surface topology. This framework successfully explains the evolution of magnetic structures across the rare-earth series \cite{jensen1991rare}. More recently, RKKY systems have regained interest in the context of frustrated magnetic materials, with the appearance of topological spin textures such as magnetic skyrmions and vortices \cite{hayami2021topological}. These textures can be stabilized either by Dzyaloshinskii-Moriya interactions and magnetic fields \cite{muhlbauer2009skyrmion} or by higher-order anisotropic terms inherent to the RKKY mechanism itself \cite{hayami2014multiple,hayami2017effective}. Moreover, the itinerant nature of the electrons in RKKY systems provides unique opportunities to probe such textures via transport phenomena, including anomalous and topological Hall effects \cite{akagi2010spin,kurumaji2019skyrmion}.

A paradigmatic example is GdRu$_2$Si$_2$, in which localized $4f$ moments of Gd ($S=7/2$) couple through conduction electrons derived primarily from Ru $4d$ orbitals \cite{nomoto2020formation}. The multiband Fermi surface revealed by quantum-oscillation measurements \cite{matsuyama2023quantum} gives rise to competing exchange interactions and complex magnetic phases. Since Gd$^{3+}$ ions carry negligible orbital moment, single-ion anisotropy is weak, implying that fourth-order RKKY processes play an essential role in stabilizing multi-$Q$ states \cite{garnier1995anisotropic,garnier1996giant}.

Although GdRu$_2$Si$_2$ has been studied for decades \cite{slaski1984magnetic}, it recently regained attention following the discovery of a skyrmion-lattice phase by resonant x-ray scattering \cite{khanh2020nanometric}. Its magnetic phase diagram includes helical states characterized by propagation vectors $\mathbf{q}_1 = (0.22\times 2\pi /a, 0, 0)$ and $\mathbf{q}_2 = (0, 0.22\times 2\pi / a, 0)$, whose superposition forms a double-$Q$ skyrmion lattice. Further studies have revealed additional topological stripe-like and high-temperature magnetic phases \cite{wood2023double,gries2025uniaxial}, highlighting the remarkable richness of this system. In particular, the magnetic structure associated to the high-temperature phase remains unresolved \cite{gries2025uniaxial}.

In this work, we employ ultrasound techniques to explore the magnetoelastic coupling in GdRu$_2$Si$_2$. We focus on the recently reported high-temperature magnetic phase \cite{gries2025uniaxial}, examining its evolution under various magnetic-field orientations. Furthermore, we analyze the sound-velocity anomalies at zero field for different acoustic modes. With this method, we perform a detailed study of the coupling between the local $4f$ moments of Gd and the itinerant $4d$ electrons of Ru, and test different ansätze for the high-temperature phase. We show, how the use of ultrasound technique can be complementary to scattering probes such as neutrons or x-rays in order to investigate complex magnetic behavior in metallic magnets.

\section{Experimental}\label{sec2}

We grew a single crystal of GdRu$_2$Si$_2$ [tetragonal ThCr$_2$Si$_2$-type structure, space group $I4/nmm$, see Fig. \hyperref[fig1]{\ref*{fig1}(a)}] using the floating-zone method, as described in more detail in Ref. \onlinecite{wood2023double}. We selected a crystal of dimensions 2.5 mm $\times$ 2.0 mm $\times$ 1.5 mm, with the longest direction of 2.5 mm along the crystallographic [110] axis.

We applied magnetic fields and controled the temperature in a Physical Property Measurement System (PPMS). We performed ultrasound experiments utilizing the transmission pulse-echo technique with phase-sensitive detection as described in Refs. \onlinecite{luthi2007physical,hauspurg2024fractionalized}. We investigated four different elastic modes shown in Fig. \hyperref[fig1]{\ref*{fig1}(a)}, for which we introduce a compact notation. The first mode is a longitudinal wave with propagation $\textbf{k}$ and polarization $\textbf{u}$ along [001], and is denoted L001. The second longitudinal mode $\textbf{k} \parallel \textbf{u} \parallel [110]$ is denoted L110. Furthermore, we studied two transverse modes, one with propagation $\textbf{k} \parallel [110]$ and polarization $\textbf{u} \parallel [001]$ denoted T001, and the second one with $\textbf{k} \parallel [110]$ and $\textbf{u} \parallel [1\bar{1}0]$ denoted T110. Thus, the two modes L001 and T001 are associated with atomic displacement along the $c$ axis, while the two other modes L110 and T110 have both propagation and polarization in the $(a,b)$ plane. We attached LiNbO$_3$ transducers (36°-Y cut and 41°-X cut for exciting longitudinal and transverse modes, respectively) to the polished surfaces of the single crystal. We used ultrasound frequencies between 60 and 90 MHz.

\section{Zero-field Results}\label{sec3}

\begin{figure}
    \centering
    \includegraphics[width=\linewidth]{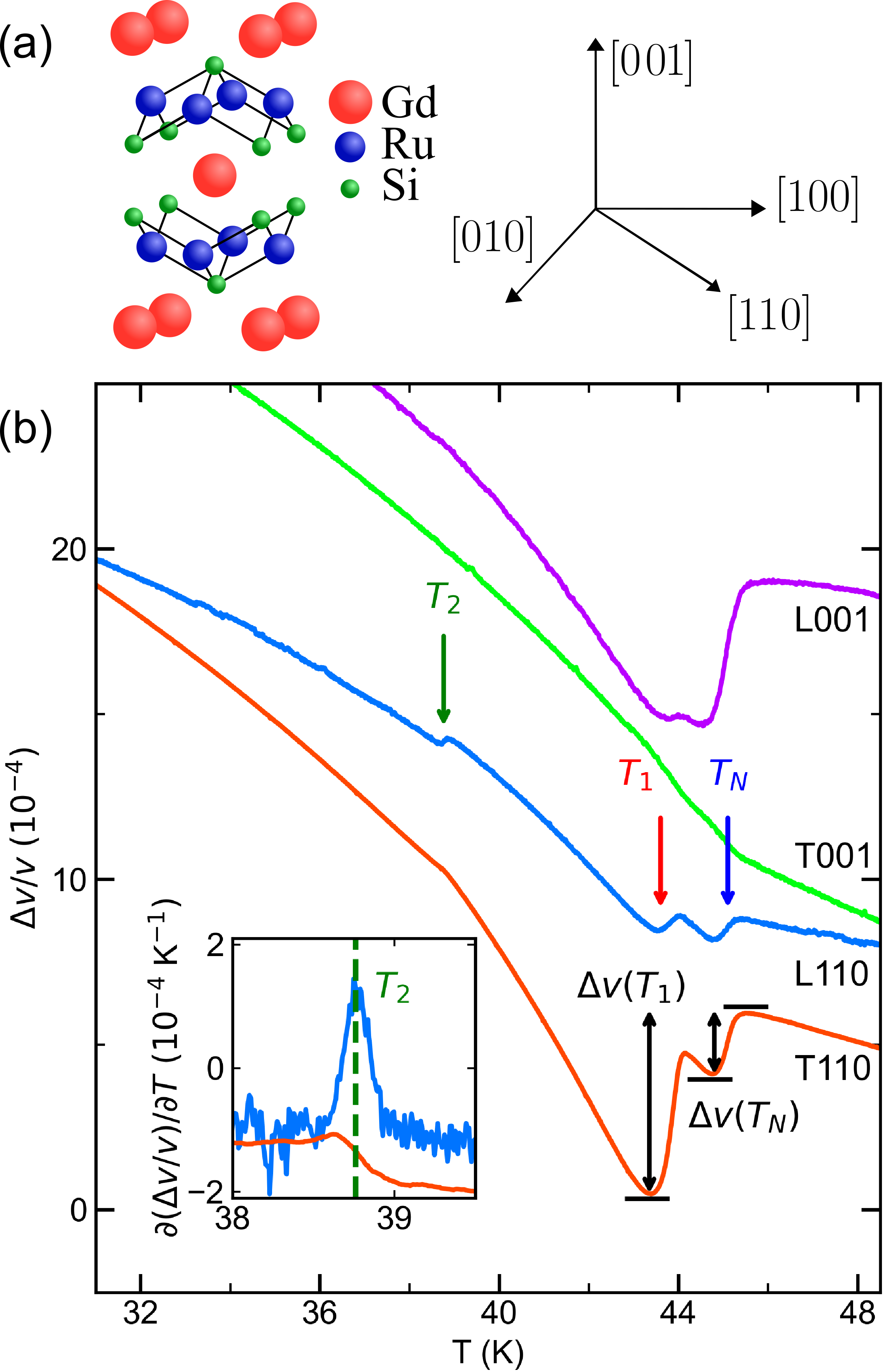}
    \caption{ (a) Crystal structure of GdRu$_2$Si$_2$. (b) Sound-velocity changes versus temperature at zero field for different elastic modes. The curves are arbitrarily shifted along $y$-axis for clarity. The insert shows the derivatives of the sound velocity changes around $T_2$ for the modes L110 and T110.}\label{fig1} 
\end{figure}

We show our results at zero field in Fig. \hyperref[fig1]{\ref*{fig1}(b)}. We observe sharp anomalies on the $\Delta v/v$ curves at the Néel transition with $T_N = 45.1$ K. We detect the strongest anomaly for the mode L001, with a jump of $\Delta v/v= -4.4\times10^{-4}$. For T110 and L110 we also observe a jump, but smaller in magnitude. Finally, for T001, we detect only a kink at this temperature. Just below $T_N$, we observe further sharp anomalies at $T_1 = 43.8$ K with a notable jump of T110, a weaker jump in L110, and rather small anomalies for L001 and T001. The narrow window between $T_1$ and $T_N$ corresponds to the high-temperature phase reported recently in \cite{gries2025uniaxial}. We determine the sound velocity anomalies at $T_i = T_N$ or $T_1$ as $\Delta v(T_i) = [v(T_i) - v_i (45.5\text{K})]/v_i(45.5\text{K})$, as shown in Fig. \hyperref[fig1]{\ref*{fig1}(b)}. We summarize the results in Table \ref{table1}, which will be discussed later within the Landau theory.

\begin{figure*}
    \centering
    \includegraphics[width=\linewidth]{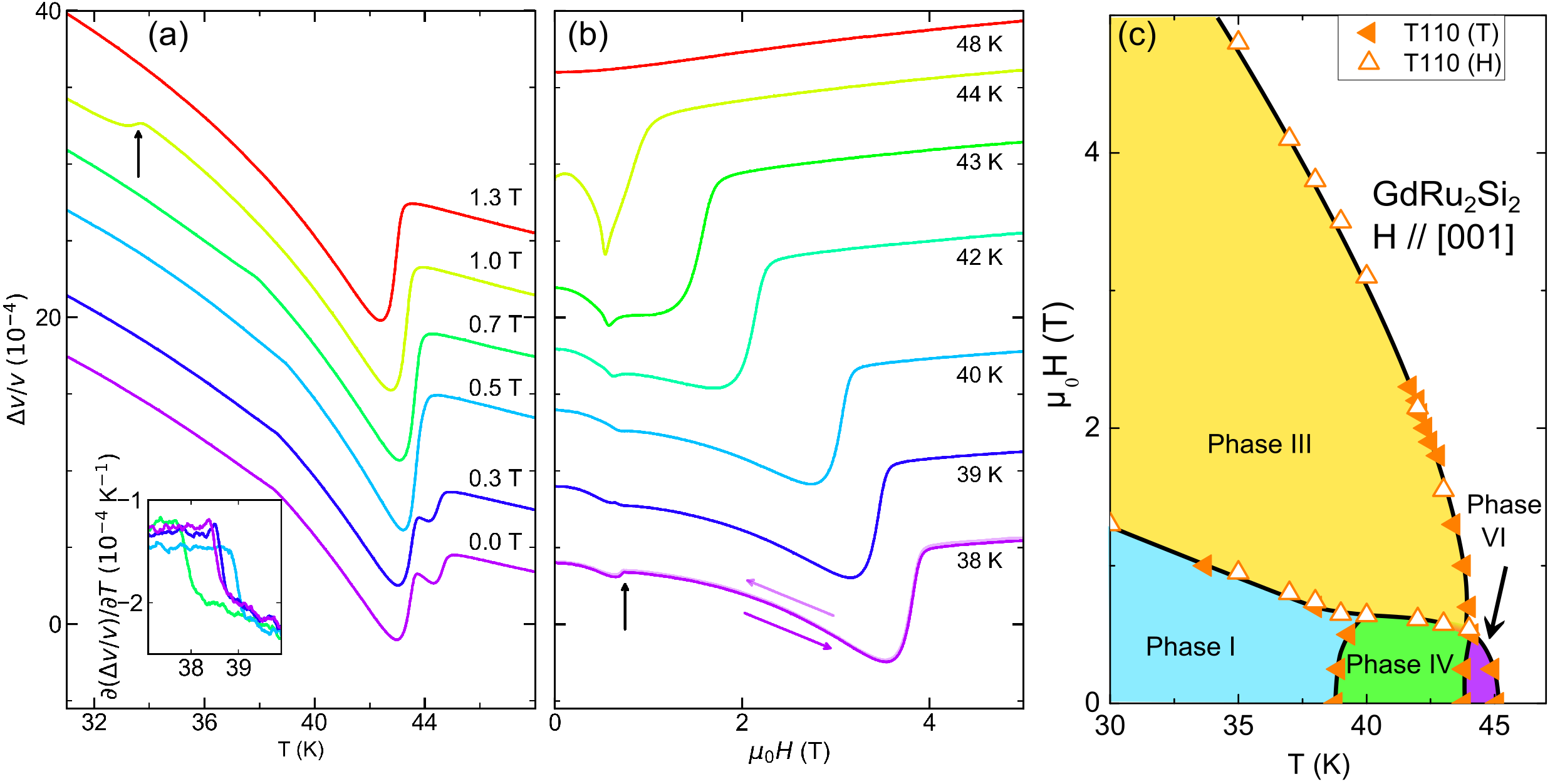}
    \caption{Sound-velocity changes of the mode T110 versus temperature for selected magnetic fields (a), and versus field for selected temperatures (b). The curves are arbitrarily shifted along $y$-axis for clarity. (c) $H-T$ phase diagram extracted from our ultrasound data, for magnetic fields applied along [001]. The black lines are guide to the eye.}\label{fig2} 
\end{figure*}

\begin{table}[h!]
    \centering
     \caption{Magnitude of the anomalies detected in the sound-velocity changes of GdRu$_2$Si$_2$ for the different elastic modes at zero field. The predicted anomalies of the different magnetic states Helical$_1$, Helical$_2$, Double$_q$, and Double$_q^*$ will be discussed in the theoretical section.}
     \label{table1}
     \scalebox{1}{\begin{tabular}{ccccccccc} 
        \toprule%
        \midrule
       $10^{-4} \times \Delta v$ & & L001 & & T001 & & L110 & & T110 \\
         \midrule
       $T_N$  & &  -4.4   & & 0   & &  -0.7   & &  -2.0    \\
       $T_1$  & &  -4.5   & & 0  & &   -1.2  & &   -6.3   \\
       \midrule
       Helical$_1$ & & -5.2 &     & 0 &     & -0.7      & &  -0.4    \\
    Helical$_2$ & & -0.04 &     & 0 &     & -0.01 &     & 0       \\
    Double$_q$  & & -5.2 &     & 0 &     & -0.6 &     & 0       \\
        Double$_q^*$ &  & -5.2 &     & 0 &     & -0.7&     & -0.1       \\
         \midrule
         \bottomrule
     \end{tabular}}
\end{table}

\

At a lower temperature, we detect another anomaly at $T_2 = 38.8$ K, which corresponds to an additional phase boundary. It is associated with a small jump in L110 and a kink in T110 (and respectively as a peak and a kind in the derivative of the sound velocities in the insert of Fig. \hyperref[fig1]{\ref*{fig1}(b)}). No clear anomaly is visible for L001 and T001 at this temperature. These results are in accordance with previous thermal-expansion observations \cite{proklevska2006magnetostriction}, where this transition does not show any anomaly along the $c$ axis but a change of slope in the thermal expansion along the $a$ axis.

\

From these zero-field results, we conclude that T110 is the most instructive: It is sensitive to all 3 transitions and shows anomalies of different magnitudes at $T_N$ and $T_1$. This allows to better distinguish the different phases. In the following, we will, thus, focus on this acoustic mode. We discuss the symmetry aspects related to the different acoustic modes the theoretical section below. 

% \begin{table}[h!]
%     \centering
%      \caption{Magnitude of the anomalies detected on GdRu$_2$Si$_2$ for the different elastic modes at zero field.}
%      \label{table1}
%      \scalebox{1}{\begin{tabular}{cccccc} 
%         \toprule%
%         \midrule
%         &  & L001 & T001 & L110 & T110 \\
%          \midrule
%          $\Delta v / v \ $($T_N$) $\times 10^{-4}$ & & -4.4 & 0.0  & -0.7 &  -2.0  \\
%          $\Delta v / v \ $($T_1$) $\times 10^{-4}$ & & -0.1 & 0.0  & -0.5 &  -4.3 \\
%          $\Delta v / v \ $($T_2$) $\times 10^{-4}$ & & 0.0& 0.0  & -0.2 &  0.0 \\
%          \midrule
%          \bottomrule
%      \end{tabular}}
% \end{table}

\section{Magnetic Fields along [001]}\label{sec3}

We show selected ultrasound data for magnetic fields applied along [001] in Fig. \hyperref[fig2]{\ref*{fig2}}. From the anomalies observed in our ultrasound data, we construct the $H-T$ phase diagram of Fig. \hyperref[fig2]{\ref*{fig2}(c)}, following the convention of Ref. \onlinecite{gries2025uniaxial} for phases I, III, IV, and VI.

\begin{figure*}
    \centering
    \includegraphics[width=.9\linewidth]{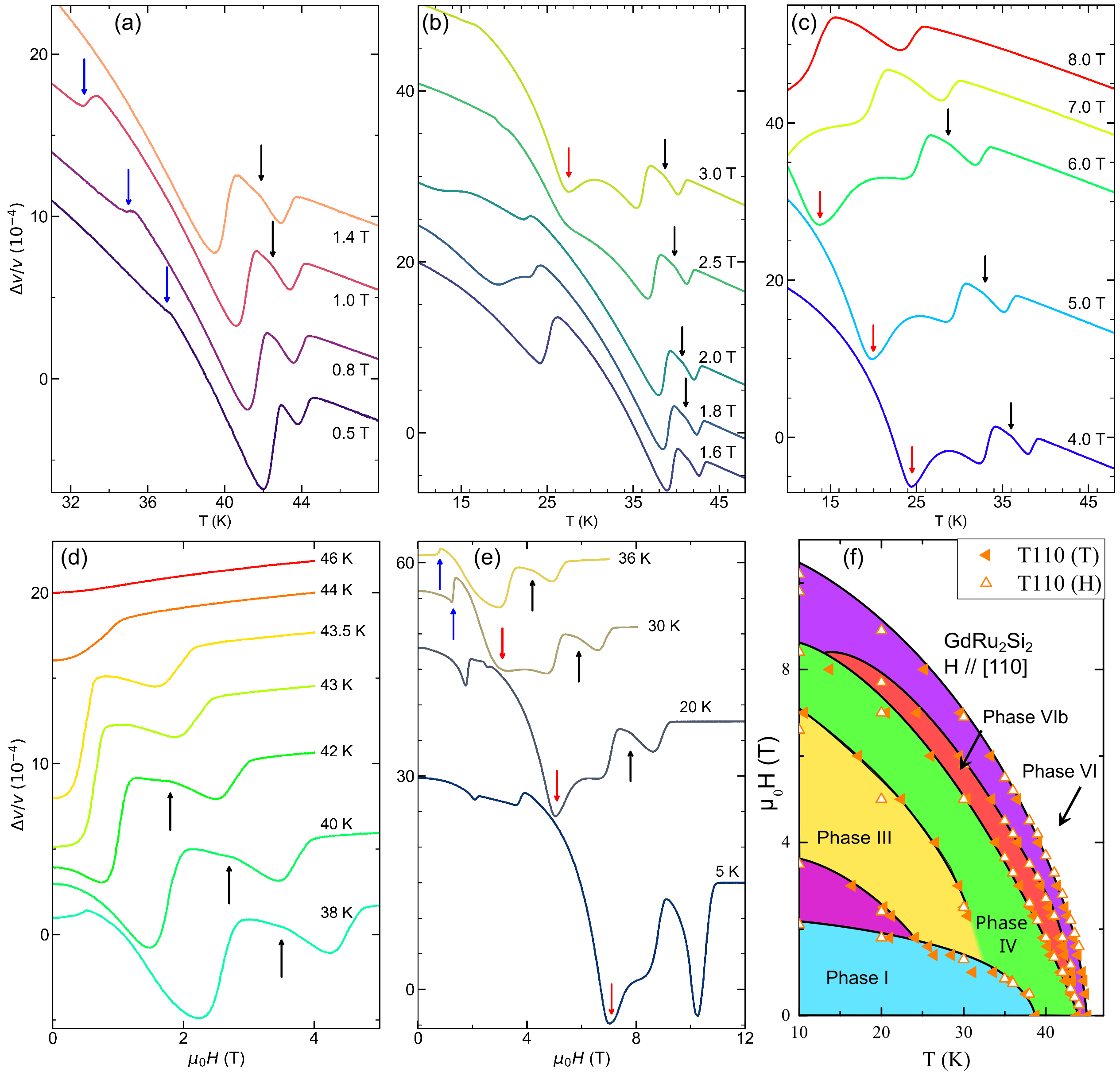}
    \caption{Sound-velocity changes of the mode T110 versus temperature for selected magnetic fields [(a), (b), and (c)], and versus field for selected temperatures [(d) and (e)]. The curves are arbitrarily shifted along $y$-axis for clarity. (f) $H-T$ phase diagram extracted from our ultrasound data, for magnetic fields applied along [110]. The black lines are guide to the eye.
    }\label{fig3} 
\end{figure*}

In the temperature sweeps [Fig. \hyperref[fig2]{\ref*{fig2}(a)}], we observe that phase VI is rapidly suppressed at a magnetic field of 0.5 T. At higher fields, the sound-velocity anomaly reflects a transition between the paramagnetic regime and phase III. We also note the evolution of the sound velocity anomaly around $T_2$. This anomaly is a kink up to 0.7 T (also visible as a jump in the derivative, shown in the insert of  Fig. \hyperref[fig2]{\ref*{fig2}(a)}), and becomes a jump at 1 T, indicated by an arrow in Fig. \hyperref[fig2]{\ref*{fig2}(a)}. While the kink anomaly is associated with the transition from phase IV to phase I, the apparition of a jump might indicate the transition to a different phase, presumably from phase III to phase I. 

% Thus, the curves at 0.5 and 0.7 T play a peculiar role: phase VI is already suppressed at high temperatures, and we observe a kink anomaly at low temperatures, suggesting the presence of a phase IV/phase I boundary. This might indicate, as we cool the temperature in this field range, to get a transition from the paramagnetic phase into phase III first, and then a crossover from phase III to phase IV. This scenario is in agreement with Ref. \onlinecite{gries2025uniaxial}, where the precise phase boundaries between phases III, IV, and VI in this region of field between 0.5 and 0.7 T could not be resolved either.

We show selected field-sweep data in Fig. \hyperref[fig2]{\ref*{fig2} (b)}. We do not observe any significant hysteresis, as shown for the 38 K curve. We detect a sharp dip at 44 K, inside phase VI. This anomaly corresponds to the transition between phase VI and the paramagnetic phase. At lower temperatures the dip becomes much smaller, and we observe a second step-like anomaly at higher fields. This indicates a transition from phase IV to phase III, followed by a transition from phase III to the paramagnetic phase. Finally, at 38 K, the dip at low fields is replaced by a jump, indicated by an arrow in Fig. \hyperref[fig2]{\ref*{fig2}(b)}. This jump signals a transition from phase I to phase III at low fields.

\section{Magnetic Fields along [110]}\label{sec3}

We show selected ultrasound data and the extracted $H-T$ phase diagram for magnetic fields applied along [110] in Fig. \hyperref[fig3]{\ref*{fig3}}. In the temperature sweeps [Figs. \hyperref[fig3]{\ref*{fig3}(a)}, \hyperref[fig3]{\ref*{fig3}(b)}, and \hyperref[fig3]{\ref*{fig3}(c)}], we observe a very rich set of anomalies. Remarkably, the anomaly associated to phase VI survives up to 9 T for this field direction. Furthermore, above 0.8 T, a kink anomaly appears inside phase VI, indicated by a black arrow in Fig. \hyperref[fig3]{\ref*{fig3}(a)}. This signals a potential intermediate phase that develops between phase VI and phase IV for this field direction, named phase VIb in our phase diagram. This kink is visible up to at least 6 T in Fig. \hyperref[fig3]{\ref*{fig3}(c)}. 

At lower temperatures, the transition between phase IV and phase I is clearly visible as a kink in the sound velocity for the data at 0.5 T, indicated by a blue arrow in Fig. \hyperref[fig3]{\ref*{fig3}(a)}. For higher fields, this kink evolves into a jump, which suggests a transition between phase I and phase III, similar to our results observed for fields along [001]. Furthermore, we do not observe any evidence of transition between phase IV and phase III in the temperature sweeps up to 2.3 T. Only above 2.3 T, there appears a minimum (around 28 K), indicated by a red arrow in Figs. \hyperref[fig3]{\ref*{fig3}(b)} and \hyperref[fig3]{(c)} and. This anomaly becomes more pronounced with increasing fields and shifts progressively towards low temperatures. It corresponds to the transition  between phase IV and phase III in our phase diagram [Fig. \hyperref[fig3]{\ref*{fig3}(f)}]. This transition line could end in a critical point around 28 K and 2.2 T. Below 2.2 T, our results indicate presumably a crossover between phase IV and phase III.

We show selected field-sweep results in Figs. \hyperref[fig3]{\ref*{fig3}(d)} and \hyperref[fig3]{\ref*{fig3}(e)}, which give us a consistent picture. At 44 K, we identify the transition into the paramagnetic regime with a jump in the sound velocity. At 43.5 K we observe two jumps, which correspond to transitions from phase IV to phase VI, followed by the transition from phase VI to the paramagnetic regime. Below 42 K, we detect the transition to phase VIb as a kink in the sound velocity, indicated by a black arrow in Fig. \hyperref[fig3]{\ref*{fig3}(d)}. We can follow this kink down to about 20 K  [Fig. \hyperref[fig3]{\ref*{fig3}(e)}]. It disappears at lower temperatures. At 38 K, we observe a kink at low fields, which indicates the transition between phase I and phase IV. At 36 K, this kink evolves into a jump [blue arrow in Fig. \hyperref[fig3]{\ref*{fig3}(e)}], which indicates the transition from phase I to phase III. At 30 K and below, we detect the transition between phase III and phase IV, indicated by red arrows in Fig. \hyperref[fig3]{\ref*{fig3}(e)}. Finally, at 20 K and below, another intermediate phase appears between 2 and 4 T, in accordance with the earlier results of Ref. \onlinecite{garnier1995anisotropic}. We summarize the $H-T$ phase diagram in Fig. \hyperref[fig3]{\ref*{fig3}(f)}. Remarkably, this phase diagram shows that phase VI is very stable for this field direction.

\section{Discussion}\label{sec3}

Our experimental results indicate a very peculiar behavior for phase VI, which is suppressed at rather small fields applied along the [001] direction, while it remains pretty stable for fields along [110]. In this section, we propose a magnetic structure for this phase, based on our data for different acoustic modes. To this aim, we develop a phenomenological Landau theory and a microscopic model for the magnetoelastic couplings, which arise from the strain dependence of the RKKY interactions in GdRu$_2$Si$_2$.

\subsection{Landau theory}

Since phase VI has a phase boundary to the paramagnetic phase, the phenomenological Landau free energy is appropriate for its description. Following the neutron-diffraction results of Refs. \onlinecite{wood2023double, paddison2024spin}, we assume that the local magnetization of the Gd atom at the site $\mathbf{r}_i$ can be written as
\begin{equation}\label{eq1}
    \begin{split}
        \mathbf{S}_i = S\sum_{\alpha}\left(\boldsymbol{\mu}_\alpha e^{i \mathbf{q}_{\alpha}\cdot\mathbf{r}_i} + \boldsymbol{\mu}_\alpha^* e^{-i\mathbf{q}_{\alpha}\cdot\mathbf{r}_i}\right),
    \end{split}{}
\end{equation}{}
where $\alpha = 1,2$ allows for double-Q states and $\boldsymbol{\mu}_\alpha$ are normalized complex vectors such that $\sum_\alpha |\boldsymbol{\mu}_\alpha|^2 = 1$. Thus, with this formulation, the value of $S$ defines the magnetic order parameter of a magnetic state characterized by the set $\{\boldsymbol{\mu}_\alpha, \mathbf{q}_{\alpha}\}$. From Refs. \onlinecite{wood2023double, paddison2024spin}, we restrict the values of $\mathbf{q}_{\alpha}$ to $\mathbf{q}_1 = (0.22\times 2\pi/a, 0, 0)$ and $\mathbf{q}_2 = (0, 0.22\times 2\pi/a, 0)$. Defining the Fourier transform $\mathbf{S}_q = \frac{1}{N}\sum_i \mathbf{S}(\mathbf{r}_i) e^{-i \mathbf{q}\cdot\mathbf{r}_i}$, we assume that the magnetic contribution to the free energy at zero field can be written as in Refs. \onlinecite{reimers1991mean,okubo2012multiple}:
\begin{equation}\label{eq2}
    \begin{split}
        \mathcal{F}_s &= \frac{1}{2}\sum_q \alpha_q \mathbf{S}_q\cdot\mathbf{S}_{-q}\\
        &+ \frac{1}{4}\sum_{q_1,q_2,q_3}\beta_{q_1,q_2,q_3}\left(\mathbf{S}_{q_1}\cdot\mathbf{S}_{q_2}\right)\left(\mathbf{S}_{q_3}\cdot\mathbf{S}_{-q_1-q_2-q_3}\right),
    \end{split}{}
\end{equation}{}
where $\alpha_q$ and $\beta_{q_1,q_2,q_3}$ are phenomenological parameters. Inserting the Eq. (\ref{eq1}), we obtain the following magnetic free energy:
\begin{equation}\label{eq2}
    \begin{split}
        &\mathcal{F}_s = \frac{A_q}{2}S^2 + \frac{B_q}{4}S^4,\\
         &A_q = \sum_{\alpha}(\alpha_{q_\alpha}+\alpha_{-q_\alpha})(\boldsymbol{\mu}_\alpha\cdot\boldsymbol{\mu}_\alpha^*)\\ 
        &B_q = 2 \sum_{\alpha\alpha'}\left[2|\boldsymbol{\mu}_\alpha|^2|\boldsymbol{\mu}_{\alpha'}|^2\beta_{q_\alpha q_\alpha q_{\alpha'}} + |\boldsymbol{\mu}_{\alpha}\cdot\boldsymbol{\mu}_{\alpha'}|^2 \beta_{q_\alpha q_{\alpha'} q_{\alpha'}}\right. \\ 
        &\ \ \ \ \ \ \ \ \ \ \ \ \ \ \ \ \ \ \ \ \ \ \ \ \ \ \ \ \ \ \ \ \ \ \ \ \ \ \ \ \left.   + \text{Re}[(\boldsymbol{\mu}_{\alpha}\cdot\boldsymbol{\mu}_{\alpha'})^2]\beta_{q_\alpha q_{\alpha'} q_{\alpha'}}\right]
    \end{split}{}
\end{equation}{}
We then include the strains $\eta_\mu$ with the bare elastic constants $C_{\mu\nu}^0$ and the magnetoelastic couplings $\lambda_\mu$, so that the total free energy can be written as
\begin{equation}\label{eq3}
    \begin{split}
        \mathcal{F}&= \mathcal{F}_s + \frac{1}{2}\sum_\mu \lambda_\mu S^2\eta_\mu + \frac{1}{2} \sum_{\mu\nu} C_{\mu\nu}^0\eta_\mu\eta_\nu.
    \end{split}{}
\end{equation}{}
In this form, our theory is similar to the Landau theory developed for the magnetoelastic coupling in CsNiCl$_3$ \cite{quirion2011magnetoelastic}. One should, in principle, include a term of the form $S^2\eta_\mu\eta_\nu$ quadratic in the order parameter and the strains. However, such coupling can only lead to a kink at the phase boundary, and we neglect such a contribution here. At the ordering temperature, the linear coupling between the strain $\eta_\mu$ and the square of the order parameter $S^2$ generates a jump of the elastic constants, which can be evaluated as
\begin{equation}\label{eq4}
    \begin{split}
        C_{\mu\nu} - C_{\mu\nu}^0 = &-\frac{\partial^2 \mathcal{F}}{\partial S \partial \eta_\mu} \left(\frac{\partial^2 \mathcal{F}}{\partial S^2}\right)^{-1}\frac{\partial^2 \mathcal{F}}{\partial \eta_\nu \partial S}
        = -\lambda_\mu \lambda_\nu \chi_S,
    \end{split}{}
\end{equation}{}
where $\chi_S = S^2 \left(\frac{\partial^2 \mathcal{F}}{\partial S^2}\right)^{-1}$. For GdRu$_2$Si$_2$, we expect that $\chi_S$ depends in a complicated way on the magnetic configuration through the terms $A_q$ and $B_q$, wich contain high-order RKKY processes as described in Refs. \onlinecite{hayami2014multiple, hayami2017effective}. However, $\chi_S$ is independent of the strain $\eta_\mu$ and we can take it as a phenomenological parameter. Thus, with this model the jumps $C_{\mu\nu} - C_{\mu\nu}^0$ depend on the strain symmetry ($\mu,\nu$) only through the magnetoelastic couplings ($\lambda_\mu$,$\lambda_\nu$). 
\subsection{Magnetoelastic coupling $\lambda_\mu$}

\begin{figure}
    \centering
    \includegraphics[width=\linewidth]{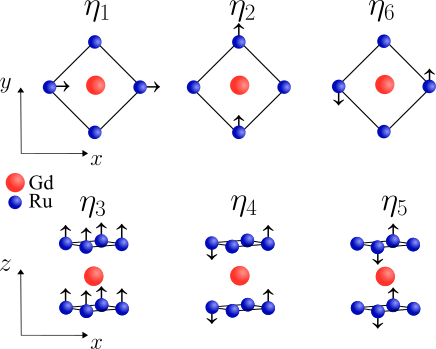}
    \caption{Visualization of the different strains on nearest neighbor Ru and Gd ions.}\label{fig5} 
\end{figure}

In this section, we evaluate the different magnetoelastic couplings $\lambda_\mu$ from a microscopic toy model. We presume that the magnetic properties of GdRu$_2$Si$_2$ are due to the RKKY interactions, which originate from the non-local Kondo coupling between Gd $4f$ and Ru $4d$ orbitals. This assumption is justified by the fact that the major contribution of the density of states at the Fermi level is dominated by Ru $4d$ bands  \cite{nomoto2020formation,bouaziz2022fermi, eremeev2023insight}. Denoting the Gd sites as $i$ and $i'$ and the Ru sites as $j$ and $j'$, we can write the RKKY Hamiltonian as
\begin{equation}\label{eq5}
    \begin{split}
       \mathcal{H}_{RKKY} &= -\sum_{ii}J_{RKKY}^{ii'}\mathbf{S}_i \cdot \mathbf{S}_{i'}, \\
       J_{RKKY}^{ii'} &= \sum_{ii'jj'}J_K^{ij}J_K^{i'j'}\chi_c(\mathbf{r}_j-\mathbf{r}_{j'}),
    \end{split}{}
\end{equation}{}
where $J_K^{ij}$ is the non-local Kondo coupling and $\chi_c(\mathbf{r})$ the spin susceptibility of the conduction electrons. In our case, we consider $J_K^{ij} = J_K$ if $i$ and $j$ are nearest neighbors as in Fig. \hyperref[fig5]{\ref*{fig5}}, and zero otherwise. The non-local Kondo coupling has been introduced in the context of the paramagnetic heavy-fermion regime \cite{weber2008heavy,sourd2024nonlocal} and we propose here to extend it in the context of the magnetic RKKY regime. Specifically, in our case, it is an essential ingredient to obtain the interactions between Gd magnetic moments and the strains. We assume an isotropic anzatz for the Kondo interaction, which decreases exponentially with increasing the Gd-Ru bond distance $J_K^{ij} = J_{K0}^{ij} \exp(-|\mathbf{r}_{ij}|/\xi)$, with $\mathbf{r}_{ij}=\mathbf{r}_{i}-\mathbf{r}_{j}$. We introduce atomic displacements $\mathbf{u}_i$ by writing $\mathbf{r}_i = \mathbf{r}_i^0 + \mathbf{u}_i$, with $|\mathbf{u}_i|$ much smaller than $a$ the lattice spacing in the $(a,b)$ plane. Then, we can expand the bond distance $r_{ij}$ and the non-local Kondo coupling $J_K^{ij}$ to first order in $|\mathbf{u}_i-\mathbf{u}_j|$, which gives
\begin{equation}\label{eq6}
    \begin{split}
        J_K^{ij} &= J_{K0}^{ij} e^{-\frac{|\mathbf{r}_{ij}^0|}{\xi}}\left(1-\sum_\mu \delta_\mu^{ij}\eta_\mu\right),
    \end{split}{}
\end{equation}{}
with the $\delta_\mu^{ij}$ given by:
\begin{equation}\label{B1}
    \begin{split}
        \delta_1^{ij} &= \frac{(r_{ij}^x)^2}{\xi r_{ij}} \  \delta_2^{ij} = \frac{(r_{ij}^y)^2}{\xi r_{ij}}, \  \delta_3^{ij} = \frac{(r_{ij}^z)^2}{\xi r_{ij}}, \\  \delta_4^{ij} &= \frac{r_{ij}^xr_{ij}^z}{\xi r_{ij}}, \ \delta_5^{ij} = \frac{r_{ij}^yr_{ij}^z}{\xi r_{ij}}, \ \delta_6^{ij} = \frac{r_{ij}^xr_{ij}^y}{\xi r_{ij}}.
    \end{split}{}
\end{equation}{}
To get the magnetoelastic coupling, we insert this expression in the RKKY Hamiltonian (\ref{eq5}) and keep only the terms linear in $\eta_\mu$. We introduce the change of variables $i' = i+z$, $j = i+z_1$, $j' = i'+ z_2$, and we obtain
\begin{equation}\label{eq7}
    \begin{split}
       \mathcal{H}_{magn-el} &= \sum_{iz\mu}\lambda_\mu^z \eta_\mu \mathbf{S}_i \cdot \mathbf{S}_{i+z},\\
       \lambda_\mu^z &= \sum_{z_1 z_2}J_{K0}^{z_1}J_{K0}^{z_2}(\delta^{z_1}_\mu+\delta^{z_2}_\mu)\chi_c(z_1-z-z_2).
    \end{split}{}
\end{equation}{}
Finally, using Eq. (\ref{eq1}), we can write the final form for the magnetoelastic couplings as
\begin{equation}\label{eq9}
    \begin{split}
        \lambda_\mu = 4\sum_{z\alpha} \lambda_\mu^z |\boldsymbol{\mu}_\alpha|^2\cos(\mathbf{q}_\alpha\cdot\mathbf{z}).
    \end{split}{}
\end{equation}{}

\subsection{Application to GdRu$_2$Si$_2$}

\begin{figure}
    \centering
    \includegraphics[width=\linewidth]{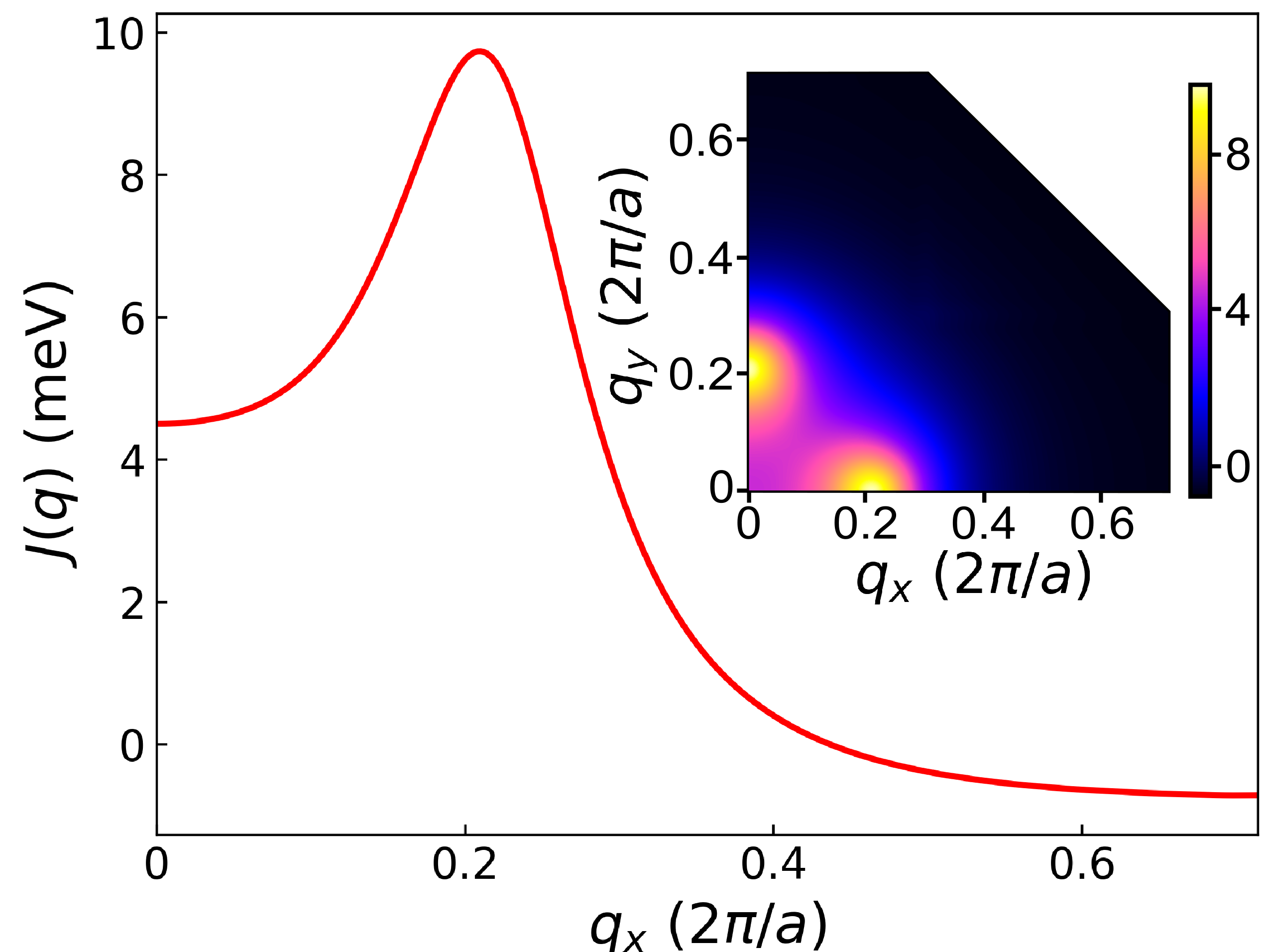}
    \caption{Fourier transform $J(q)$ of the magnetic interactions up to 60 unit cells. The inset shows a colormap of $J(q)$ on the $q_z = 0$ plane of the Brillouin zone. }\label{fig6} 
\end{figure}

In order to evaluate $\lambda_\mu^z$ from Eq. (\ref{eq7}), we take the simple ansatz of an Ornstein-Zernike form for the conduction-electron spin susceptibility \cite{eremin2001large,abanov2003quantum}, which reproduces the maxima coming from the nesting properties of the Fermi surface \cite{bouaziz2022fermi}:
\begin{equation}\label{eq10}
    \begin{split}
        \chi_c(z) &= \frac{1}{N}\sum_q e^{iqz}\chi_q, \\ 
        \chi_q &= \chi_0\sum_{q_\alpha = \pm q_1, \pm q_2} \frac{ \zeta^{-2}}{\zeta^{-2} + (q-q_\alpha)^2}.
    \end{split}{}
\end{equation}{}
In this expression, the factor $\zeta$ can be seen as a coherence length, which controls the long-distance decay of the RKKY interactions. We choose $\zeta = 2 a $, which generates a sufficiently fast decay that allows us to restrict the sum over $z$ in Eq. (\ref{eq9}) to the first 60 unit cells. Furthermore, we fix $J_K^2\chi_0 = 6$ $\mu$eV. With this choice, the Fourier transform of our magnetic coupling $J_{RKKY}^{ii'}$ shown in Fig. \hyperref[fig6]{\ref*{fig6}} reproduce qualitatively the results from ab-initio calculations of Ref. \onlinecite{bouaziz2022fermi}. This $J(q)$ is also in qualitative agreement with the recent inelastic neutron scattering results of Ref. \onlinecite{wood2025magnon}.

% \

% , this value of $\zeta$ reproduces qualitatively the magnetic couplings obtained from ab-initio approaches in Ref. \onlinecite{bouaziz2022fermi}, as shown in Table \ref{table2}. Apart from the coupling $J_3$, we find a good agreement between our simple theory and the  ab-initio results \cite{bouaziz2022fermi}. For the couplings $J_5$ and $J_7$, we have a better agreement with the ab-initio couplings compared to the $J_{1-8}$ model where a restricted number of coupling $J$ was used to fit the magnon dispersion \cite{wood2025magnon}.

% \begin{table}[h!]
%     \centering
%      \caption{Comparison between the first 8 RKKY couplings for the model presented here with  $J_K^2\xi_0 = 8$ $\mu$eV and $\zeta = 2 a $, and the ab-initio results of Ref \cite{bouaziz2022fermi}, as well as the $J_{1-8}$ model of Ref \cite{wood2025magnon}.}
%      \label{table2}
%      \scalebox{1}{\begin{tabular}{cccccccccc} 
%         \toprule%
%         \midrule
%         &  & $J_1$ & $J_2$ & $J_3$ & $J_4$ & $J_5$ & $J_6$ & $J_7$ & $J_8$  \\
%          \midrule
%         Eq. (\ref{eq5}) ($\mu$eV) & & -43.1 & -43.4  & -15.5 &  -7.2 & -2.4 & 3.7 & -20.5 & 4.2  \\
%          Ab-initio \cite{bouaziz2022fermi} ($\mu$eV) & & ins & ide  & Ref & \cite{wood2025magnon} & - & - & - & - \\
%         $J_{1-8}$  \cite{wood2025magnon} ($\mu$eV) & & -33.9 & -65.1  & 13.6 &  -1.1 & 10.8 & 6.0 & 0 & 3.0 \\
%          \midrule
%          \bottomrule
%      \end{tabular}}
% \end{table}
\

We considered the different magnetic states proposed for phase III and phase IV in Refs. \onlinecite{khanh2020nanometric, paddison2024spin, wood2023double, yasui2020imaging}. From Eq. (\ref{eq7}), we obtain the same contribution to the magnetoelastic energy from helical and amplitude-modulated phases. Thus, our method does not allow us to distinguish between them. Furthermore, from Eq. (\ref{eq7}), we also notice that the direction of the magnetic moment is arbitrary. Thus, we can restrict ourselves to two different states: The single-Q helical state with wave vector $\mathbf{q}_1$ denoted Helical$_1$, and the double-Q state with wave vectors  $\mathbf{q}_1$ and $\mathbf{q}_2$ denoted Double$_q$. Note that the Helical$_1$ state corresponds to phase IV \cite{paddison2024spin}.  We added a second helical state with wave vector $\mathbf{q}_1 + \mathbf{q}_2$ denoted Helical$_2$ for comparison, as well as an asymmetric double-Q state with unequal contributions of $\mathbf{q}_1$ and $\mathbf{q}_2$, denoted Double$_{q}^*$.

\ 

\begin{table}[h!]
    \centering
     \caption{Different magnetic phases defined by $\{\boldsymbol{\mu}_\alpha, \mathbf{q}_{\alpha}\}$, and the associated magnitude of the magnetoelastic couplings.}
     \label{table2}
     \scalebox{1}{\begin{tabular}{cccccc} 
        \toprule%
        \midrule
        &   & Helical$_1$ & Helical$_2$ & Double$_{q}$ & Double$_{q}^*$ \\
         \midrule
         $\boldsymbol{\mu}_1$ & & $\frac{1}{\sqrt{2}}(1,i,0)$  & $\frac{1}{\sqrt{2}}(1,i,0)$ &  $\frac{1}{\sqrt{2}}(1,0,0)$ &  $\frac{1}{2}(1,0,0)$ \\
         $\mathbf{q}$ &  & $\mathbf{q}_1$  & $\mathbf{q}_1 + \mathbf{q}_2$ &  $\mathbf{q}_1$ &  $\mathbf{q}_1$ \\
         $\boldsymbol{\mu}_2$ & & -  & - &  $\frac{1}{\sqrt{2}}(0,1,0)$ &  $\frac{\sqrt{3}}{2}(0,1,0)$ \\
         $\mathbf{q}'$ & & - & -  & $\mathbf{q}_2$ &  $\mathbf{q}_2$ \\
         \midrule
         $\Lambda_1$ (meV$\textup{\AA}$) & & 0.7 & 0.1 & 0.7 &  0.7 \\
         $\Lambda_2$ (meV$\textup{\AA}$)& & 0.5 & 0.1  & 0.7 &  0.6 \\
         $\Lambda_3$ (meV$\textup{\AA}$)& & 1.7 & 0.2  & 1.7 &  1.7 \\
         \midrule
         \bottomrule
     \end{tabular}}
\end{table}

Upon evaluating the magnetoelastic couplings, we find that only the compression strains $\eta_1$, $\eta_2$, and $\eta_3$ contribute significantly. This is a consequence of our assumption for the strain dependence of the Kondo interaction in Eq. (\ref{eq5}), which contributes only to compression and stretching of the Gd-Ru bonds. Defining $\Lambda_\mu = \lambda_\mu \xi$, we summarize our results for the different states in Table \ref{table2}. The magnetoelastic couplings $\Lambda_4$, $\Lambda_5$, and $\Lambda_6$ are negligible. The relative values between the $\Lambda_\mu$ will be directly related to the size of the jumps in the sound velocity between each mode.

\ 

For all states, we found that the coupling $\Lambda_3$ dominates strongly. This becomes clear from Fig. \hyperref[fig5]{\ref*{fig5}}, where the strain $\eta_3$ gives the largest number of Ru atoms contributing to the magnetoelastic coupling. The couplings $\Lambda_1$ and $\Lambda_2$ are equal in the Helical$_2$ and Double$_q$ states, as expected by symmetry between the [100] and the [010] axes. In the Helical$_1$ and in the Double$_q^*$ states, however, this symmetry is broken, and the degeneracy between $\Lambda_1$ and $\Lambda_2$  is lifted. Furthermore, we can distinguish the Helical$_2$ state by comparing the value of $\Lambda_3$, which is much smaller for the Helical$_2$ state, while it has the same value in the other 3 states.

\

From the relative values of the $\Lambda_\mu$, we obtain directly the relative values of the jumps in the elastic constants thanks to Eq. (\ref{eq4}). We evaluate the corresponding sound velocity jumps using $\rho v_{L001}^2 = C_{33}$, $\rho v_{T001}^2 = C_{44}$, $2\rho v_{L110}^2 = C_{11} + C_{12} + 2C_{66}$, and $2\rho v_{T110}^2 = C_{11} - C_{12}$. We obtain

\begin{equation}\label{C1}
    \begin{split}
         \Delta v_{L001} &= -K_{L001} \Lambda_3^2 \\
        \Delta v_{T001} &= 0 \\
        \Delta v_{L110} &= - K_{L110}\frac{\Lambda_1^2 + \Lambda_1\Lambda_2}{2}\\
        \Delta v_{T110} &= -K_{T110}\frac{\Lambda_1^2 - \Lambda_1\Lambda_2}{2}.
    \end{split}{}
\end{equation}{}
where $K_i =  \chi_S / 2 \rho \xi^2 v_i^2 = K / v_i^2$. From this Eq. (\ref{C1}), we conclude that no jump should be observed in the T001 mode, in accordance with our experimental data. We also see that $\Delta v_{T110} = 0$ if $\Lambda_1 = \Lambda_2$, i.e. when the  [100] and [010] axes are equivalent, such as for the Helical$_2$ and Double$_q$ states. In order to further compare our predictions with the experimental data, we fix the phenomenological parameter to $K = 3\times 10^{10}$ (meVs)$^{-2}$, and we use $v_{L001} = 4200$ m/s, $v_{T001} = 2800$ m/s, $v_{L110} = 4600$ m/s, and $v_{T110} = 2600$ m/s measured at $T = 50$ K. The resulting anomalies for the different magnetic states are given in Table \ref{table1}.
 
\ 

We propose that the transition into the Helical$_1$ state at $T = T_1$ can also be described within the Landau theory, since $T_1$ is very close to $T_N$. As shown in Table \ref{table1}, our prediction for the Helical$_1$ state matches well the anomalies observed at $T_1$, with a strong anomaly for the mode L001, no anomaly for T001, and a weak anomaly for L110. Only the anomaly for T110 is underestimated one order of magnitude. Nevertheless, our toy model shows a good qualitative agreement with the ultrasound results for the anomalies at $T_1$.

% \begin{table}[h!]
%     \centering
%      \caption{Magnitude of the magnetoelastic couplings for the different phases in GdRu$_2$Si$_2$.}
%      \label{table3}
%      \scalebox{1}{\begin{tabular}{ccccccccc} 
%         \toprule%
%         \midrule
%         & & L001 & & T001 & & L110 & & T110 \\
%          \midrule
%        $v_i$ (m/s)  & &  4200   & &   2800  & &  4600   & & 2600     \\
%        $10^4\frac{\Delta v_i}{v_i}$ ($T_N$)  & &  -4.4   & & 0   & &  -0.7   & &  -2.0    \\
%        $10^4\frac{\Delta v_i}{v_i}$ ($T_1$)  & &  -4.5   & & 0  & &   -1.2  & &   -6.3   \\
%        \midrule
%        $10^4\frac{\Delta v_i}{v_i}$ (Helical$_1$) & & -5.2 &     & 0 &     & -0.7      & &  -0.4    \\
%        $10^4\frac{\Delta v_i}{v_i}$ (Helical$_2$) & & -0.04 &     & 0 &     & -0.01 &     & 0       \\
%        $10^4\frac{\Delta v_i}{v_i}$ (Double$_q$)  & & -5.2 &     & 0 &     & -0.6 &     & 0       \\
%        $10^4\frac{\Delta v_i}{v_i}$ (Double$_q^*$) &  & -5.2 &     & 0 &     & -0.7&     & -0.1       \\
%          \midrule
%          \bottomrule
%      \end{tabular}}
% \end{table}

Finally, we compare our prediction for $T_N$, associated with the unknown phase VI. Both the Helical$_1$ and Double$_q$ configurations give comparable jumps in L001, which matches our experimental data. At $T_N$, the anomaly in T110 is much weaker than at $T_1$ but yet is finite, which implyes a partial symmetry breaking between [100] and [010]. Thus, phase VI might be an intermediate between the Helical$_1$ state at $T_1$ which maximally breaks this symmetry, and the Double$_q$ state which do not break this symmetry and thus gives no jump for T110. This leads us to propose that phase VI corresponds to an intermediate Double$_q^*$ state, characterized by unequal amplitudes of the two propagation vectors $\mathbf{q}_1$ and $\mathbf{q}_2$. In our case we used $|\boldsymbol{\mu}_2|^2 = 3 |\boldsymbol{\mu}_1|^2$ as shown in Table  \ref{table2}.

\section{Conclusion}

We investigated the low-temperature magnetoacoustic properties of single-crystalline GdRu$_2$Si$_2$ and identified a series of sharp anomalies in the sound velocity, enabling the construction of detailed magnetic phase diagrams for fields applied along the $[001]$ and $[110]$ directions.

We gave particular attention to the recently reported high-temperature phase VI. We find that this phase exhibits strong directional sensitivity: It is suppressed by modest fields of 0.5 T applied along $[001]$, yet remains stable up to 9 T along $[110]$. Measurements of multiple acoustic modes at zero field further reveal distinctive magnetoelastic signatures associated with this phase.

To interpret these findings, we developed a Landau framework and a microscopic toy model describing the strain dependence of the RKKY interactions in GdRu$_2$Si$_2$. The model reproduces the main experimental trends and points to an asymmetric double-$Q$ magnetic structure for phase VI.

These results highlight the intimate coupling between lattice strain and itinerant-electron-mediated magnetism in GdRu$_2$Si$_2$, and demonstrate how ultrasound experiments can serve as a sensitive probe of multi-$Q$ magnetic textures in metallic magnets.

% In this work, we have investigated the low temperature magnetoacoustic properties of a GdRu$2$Si$_2$ single crystal. We have detected a rich set of jump and kink anomalies in the sound velocity, which allows us to construct detailed phase diagrams for magnetic fields applied along $[001]$ and $[110]$ directions.

% In particular, our study focuses on the recently reported high temperature phase VI. We show that this phase is very sensitive to the magnetic field direction, being suppressed at 0.5 T for fields along $[001]$, while it remains stable up to 8 T for fields along $[110]$. Furthermore, we analyzed different acoustic modes at zero field in order to explore in more details the magnetoelastic properties of this phase. We propose a Landau theory and a microscopic toy model for strain dependence of the RKKY interactions in GdRu$_2$Si$_2$. Our model reproduce qualitatively the experimental results, and suggests an asymmetric double-Q structure for phase VI. 

\section{Aknowledgments}
We acknowledge support from the Deutsche Forschungsgemeinschaft (DFG)
through SFB 1143 (Project No.\ 247310070) and the
W\"{u}rzburg-Dresden Cluster of Excellence on Complexity and 
Topology in Quantum Matter--$ct.qmat$ (EXC 2147, Project No.\ 390858490),
as well as the support of the HLD at HZDR, member of the European
Magnetic Field Laboratory (EMFL). We further acknowledge support
under the European Union’s Horizon 2020 research and innovation
programme through the ISABEL project (No. 871106). The work at the University of Warwick was financially supported by two Engineering and Physical Sciences Research Council grants: Grant No. EP/T005963/1 and the U.K. Skyrmion Project Grant No. EP/N032128/1

\bibliography{biblio_JS_GdRu2Si2.bib}

@article{brando2016metallic,
  title={Metallic quantum ferromagnets},
  author={Brando, M. and Belitz, D. and Grosche, F. M. and Kirkpatrick, T.R.},
  journal={Rev. Mod. Phys.},
  volume={88},
  number={2},
  pages={025006},
  year={2016},
  publisher={APS}
}

@article{szilva2023quantitative,
  title={Quantitative theory of magnetic interactions in solids},
  author={Szilva, A. and Kvashnin, Y. and Stepanov, E.A. and Nordstr{\"o}m, L. and Eriksson, O. and Lichtenstein, A.I. and Katsnelson, M.I.},
  journal={Rev. Mod. Phys.},
  volume={95},
  number={3},
  pages={035004},
  year={2023},
  publisher={APS}
}

@article{sourd2024nonlocal,
  title={Nonlocal {K}ondo coupling and selective doping from cerium f electrons in iron-based superconductors},
  author={Sourd, J. and Tenc{\'e}, S. and Gaudin, E. and Burdin, S.},
  journal={Phys. Rev. B},
  volume={109},
  number={4},
  pages={045117},
  year={2024},
  publisher={APS}
}

@article{eremeev2023insight,
  title={Insight into the electronic structure of the centrosymmetric skyrmion magnet {G}d{R}u$_2${S}i$_2$},
  author={Eremeev, S.V. and Glazkova, D. and Poelchen, G. and Kraiker, A. and Ali, K. and Tarasov, A.V. and Schulz, S. and Kliemt, K. and Chulkov, E.V. and Stolyarov, V.S. and others},
  journal={Nano. Adv.},
  volume={5},
  number={23},
  pages={6678--6687},
  year={2023},
  publisher={Royal Society of Chemistry}
}

@article{bouaziz2022fermi,
  title={Fermi-surface origin of skyrmion lattices in centrosymmetric rare-earth intermetallics},
  author={Bouaziz, J. and Mendive-Tapia, E. and Bl{\"u}gel, S. and Staunton, J.B.},
  journal={Physical review letters},
  volume={128},
  number={15},
  pages={157206},
  year={2022},
  publisher={APS}
}

@article{ruderman1954indirect,
  title={Indirect exchange coupling of nuclear magnetic moments by conduction electrons},
  author={Ruderman, M.A. and Kittel, C.},
  journal={Phys. Rev.},
  volume={96},
  number={1},
  pages={99},
  year={1954},
  publisher={APS}
}

@article{kasuya1956theory,
  title={A theory of metallic ferro-and antiferromagnetism on {Z}ener's model},
  author={Kasuya, T.},
  journal={Prog. Theor. Phys.},
  volume={16},
  number={1},
  pages={45--57},
  year={1956},
  publisher={Oxford University Press}
}

@article{yosida1957magnetic,
  title={Magnetic properties of {C}u-{M}n alloys},
  author={Yosida, K.},
  journal={Phys. Rev.},
  volume={106},
  number={5},
  pages={893},
  year={1957},
  publisher={APS}
}

@book{jensen1991rare,
  title={Rare-earth magnetism: structures and excitations},
  author={Jensen, J. and Mackintosh, A.R.},
  year={1991},
  publisher={Oxford University Press}
}

@article{hayami2021topological,
  title={Topological spin crystals by itinerant frustration},
  author={Hayami, S. and Motome, Y.},
  journal={J. Phys. Condens. Matter},
  volume={33},
  number={44},
  pages={443001},
  year={2021},
  publisher={IOP Publishing}
}

@article{eremin2001large,
  title={Large magnetoresistance and critical spin fluctuations in {G}d{I}$_2$},
  author={Eremin, I. and Thalmeier, P. and Fulde, P. and Kremer, R.K. and Ahn, K. and Simon, A.},
  journal={Physical Review B},
  volume={64},
  number={6},
  pages={064425},
  year={2001},
  publisher={APS}
}

@article{abanov2003quantum,
  title={Quantum-critical theory of the spin-fermion model and its application to cuprates: Normal state analysis},
  author={Abanov, A. and Chubukov, A.V. and Schmalian, J.},
  journal={Advances in Physics},
  volume={52},
  number={3},
  pages={119--218},
  year={2003},
  publisher={Taylor \& Francis}
}

@article{muhlbauer2009skyrmion,
  title={Skyrmion lattice in a chiral magnet},
  author={Muhlbauer, S. and Binz, B. and Jonietz, F. and Pfleiderer, C. and Rosch, A. and Neubauer, A. and Georgii, R. and Boni, P.},
  journal={Science},
  volume={323},
  number={5916},
  pages={915--919},
  year={2009},
  publisher={American Association for the Advancement of Science}
}

@article{hayami2014multiple,
  title={Multiple-{Q} instability by (d- 2)-dimensional connections of {F}ermi surfaces},
  author={Hayami, S. and Motome, Y.},
  journal={Phys. Rev. B},
  volume={90},
  number={6},
  pages={060402},
  year={2014},
  publisher={APS}
}

@article{hayami2017effective,
  title={Effective bilinear-biquadratic model for noncoplanar ordering in itinerant magnets},
  author={Hayami, S. and Ozawa, R. and Motome, Y.},
  journal={Phys. Rev. B},
  volume={95},
  number={22},
  pages={224424},
  year={2017},
  publisher={APS}
}

@article{akagi2010spin,
  title={Spin chirality ordering and anomalous {H}all effect in the ferromagnetic {K}ondo lattice model on a triangular lattice},
  author={Akagi, Y. and Motome, Y.},
  journal={J. Phys. Soc. Japan},
  volume={79},
  number={8},
  pages={083711},
  year={2010},
  publisher={The Physical Society of Japan}
}

@article{kurumaji2019skyrmion,
  title={Skyrmion lattice with a giant topological {H}all effect in a frustrated triangular-lattice magnet},
  author={Kurumaji, T. and Nakajima, T. and Hirschberger, M. and Kikkawa, A. and Yamasaki, Y. and Sagayama, H. and Nakao, H. and Taguchi, Y. and Arima, T. and Tokura, Y.},
  journal={Science},
  volume={365},
  number={6456},
  pages={914--918},
  year={2019},
  publisher={American Association for the Advancement of Science}
}

@article{slaski1984magnetic,
  title={Magnetic properties of {$RE$}{R}u$_2$Si$_2$ ({$RE$}= {P}r, {N}d, {G}d, {T}b, {D}y, {E}r) interm etallics},
  author={{\'S}laski, M. and Szytu{\l}a, A. and Leciejewicz, J. and Zygmunt, A.},
  journal={J. Magn. Magn. Mat.},
  volume={46},
  number={1-2},
  pages={114--122},
  year={1984},
  publisher={Elsevier}
}

@article{nomoto2020formation,
  title={Formation mechanism of the helical {Q} structure in {G}d-based skyrmion materials},
  author={Nomoto, T. and Koretsune, T. and Arita, R.},
  journal={Phys. Rev. Lett.},
  volume={125},
  number={11},
  pages={117204},
  year={2020},
  publisher={APS}
}

@article{matsuyama2023quantum,
  title={Quantum oscillations in the centrosymmetric skyrmion-hosting magnet {G}d{R}u$_2${S}i$_2$},
  author={Matsuyama, N. and Nomura, T. and Imajo, S. and Nomoto, T. and Arita, R. and Sudo, K. and Kimata, M. and Khanh, N.D. and Takagi, R. and Tokura, Y. and others},
  journal={Phys. Rev. B},
  volume={107},
  number={10},
  pages={104421},
  year={2023},
  publisher={APS}
}

@article{khanh2020nanometric,
  title={Nanometric square skyrmion lattice in a centrosymmetric tetragonal magnet},
  author={Khanh, N.D. and Nakajima, T. and Yu, X. and Gao, S. and Shibata, K. and Hirschberger, M. and Yamasaki, Y. and Sagayama, H. and Nakao, H. and Peng, L. and others},
  journal={Nat. Nano.},
  volume={15},
  number={6},
  pages={444--449},
  year={2020},
  publisher={Nature Publishing Group UK London}
}

@article{hauspurg2024fractionalized,
  title={Fractionalized excitations probed by ultrasound},
  author={Hauspurg, A. and Zherlitsyn, S. and Helm, T. and Felea, V. and Wosnitza, J. and Tsurkan, V. and Choi, K.Y. and Do, S.H. and Ye, M. and Brenig, W. and others},
  journal={Phys. Rev. B},
  volume={109},
  number={14},
  pages={144415},
  year={2024},
  publisher={APS}
}

@article{garnier1995anisotropic,
  title={Anisotropic metamagnetism in {G}d{R}u$_2${S}i$_2$},
  author={Garnier, A. and Gignoux, D. and Iwata, N. and Schmitt, D. and Shigeoka, T. and Zhang, F.Y.},
  journal={J. Magn. Magn. Mat.},
  volume={140},
  pages={899--900},
  year={1995},
  publisher={Elsevier}
}

@article{garnier1996giant,
  title={Giant magnetic anisotropy in tetragonal {G}d{R}u$_2${G}e$_2$ and {G}d{R}u$_2${S}i$_2$},
  author={Garnier, A. and Gignoux, D. and Schmitt, D. and Shigeoka, T.},
  journal={Physica B},
  volume={222},
  number={1-3},
  pages={80--86},
  year={1996},
  publisher={Elsevier}
}

@article{yasui2020imaging,
  title={Imaging the coupling between itinerant electrons and localised moments in the centrosymmetric skyrmion magnet {G}d{R}u$_2${S}i$_2$},
  author={Yasui, Y. and Butler, C.J. and Khanh, N.D. and Hayami, S. and Nomoto, T. and Hanaguri, T. and Motome, Y. and Arita, R. and Arima, T. and Tokura, Y. and others},
  journal={Nat. Comm.},
  volume={11},
  number={1},
  pages={5925},
  year={2020},
  publisher={Nature Publishing Group UK London}
}

@article{proklevska2006magnetostriction,
  title={Magnetostriction measurement of {G}d{R}u$_2${S}i$_2$ single crystal},
  author={Prokle{\v{s}}ka, J. and Vejpravov{\'a}, J. and Sechovsk{\`y}, V.},
  journal={J. Phys. Conf. Ser.},
  volume={51},
  number={1},
  pages={127},
  year={2006},
  organization={IOP Publishing}
}

@article{wood2023double,
  title={Double-{Q} ground state with topological charge stripes in the centrosymmetric skyrmion candidate {G}d{R}u$_2${S}i$_2$},
  author={Wood, G.D.A. and Khalyavin, D.D. and Mayoh, D.A. and Bouaziz, J. and Hall, A.E. and Holt, S.J.R. and Orlandi, F. and Manuel, P. and Bl{\"u}gel, S. and others},
  journal={Phys. Rev. B},
  volume={107},
  number={18},
  pages={L180402},
  year={2023},
  publisher={APS}
}

@article{paddison2024spin,
  title={Spin dynamics of the centrosymmetric skyrmion material {G}d{R}u$_2${S}i$_2$},
  author={Paddison, J.A.M. and Bouaziz, J. and May, A.F. and Zhang, Q. and Calder, S. and Abernathy, D. and Staunton, J.B. and Bl{\"u}gel, S. and Christianson, A.D.},
  journal={Cell Rep. Phys. Sci.},
  volume={5},
  number={11},
  year={2024},
  publisher={Elsevier}
}

@article{reimers1991mean,
  title={Mean-field approach to magnetic ordering in highly frustrated pyrochlores},
  author={Reimers, J.N. and Berlinsky, A.J. and Shi, A.C.},
  journal={Phys. Rev. B},
  volume={43},
  number={1},
  pages={865},
  year={1991},
  publisher={APS}
}

@article{okubo2012multiple,
  title={Multiple-q States and the Skyrmion Lattice of the Triangular-Lattice Heisenberg<? format?> Antiferromagnet under Magnetic Fields},
  author={Okubo, T. and Chung, S. and Kawamura, H.},
  journal={Phys. Rev. Lett.},
  volume={108},
  number={1},
  pages={017206},
  year={2012},
  publisher={APS}
}

@article{quirion2011magnetoelastic,
  title={Magnetoelastic coupling within a Landau model of phase transitions: Application to the frustrated triangular antiferromagnet {C}s{N}i{C}l$_3$},
  author={Quirion, G. and Han, X. and Plumer, M.L.},
  journal={Phys. Rev. B},
  volume={84},
  number={1},
  pages={014408},
  year={2011},
  publisher={APS}
}

@article{gries2025uniaxial,
  title={Uniaxial pressure effects, phase diagram, and tricritical point in the centrosymmetric skyrmion lattice magnet {G}d{R}u$_2${S}i$_2$},
  author={Gries, L. and Kleinbeck, T. and Mayoh, D.A. and Wood, G.D.A. and Balakrishnan, G. and Klingeler, R.},
  journal={Phys. Rev. B.},
  volume={111},
  number={6},
  pages={064419},
  year={2025},
  publisher={APS}
}

@article{wood2025magnon,
  title={A magnon band analysis of {G}d{R}u$_2${S}i$_2$ in the field-polarized state},
  author={Wood, G.D.A. and Stewart, J.R. and Mayoh, D.A. and Paddison, J.A.M. and Bouaziz, J. and Tobin, S.M. and Petrenko, O.A. and Lees, M.R. and Manuel, P. and Staunton, J.B. and others},
  journal={npj Quantum Materials},
  volume={10},
  number={1},
  pages={39},
  year={2025},
  publisher={Nature Publishing Group UK London}
}

@book{luthi2007physical,
  title={Physical {A}coustics in the {S}olid {S}tate},
  author={L{\"u}thi, B.},
  volume={148},
  year={2006},
  publisher={Springer, Berlin, Heidelberg}
}

@article{weber2008heavy,
  title={Heavy-fermion metals with hybridization nodes: Unconventional {F}ermi liquids and competing phases},
  author={Weber, H. and Vojta, M.},
  journal={Phys. Rev. B},
  volume={77},
  number={12},
  pages={125118},
  year={2008},
  publisher={APS}
}

\end{document}